\documentstyle[12pt,leqno]{article}
\newtheorem{theorem}{Theorem}[section]
\newtheorem{definition}[theorem]{Definition}

\newtheorem{example}[theorem]{Example}
\newtheorem{lemma}[theorem]{Lemma}
\newtheorem{proposition}[theorem]{Proposition}
\newtheorem{remark}[theorem]{Remark}

\catcode`\@=11
\def\@begintheorem#1#2{\it \trivlist \item[\hskip
 \labelsep{\bf #1\ #2.}]}
\def\@opargbegintheorem#1#2#3{\it \trivlist\item[\hskip%
 \labelsep{\bf #1\ #2.\ (#3)}]}
\def\@endtheorem{\endtrivlist}

\def\@listI{\leftmargin\leftmargini \parsep 1pt plus 2.5pt
 minus 1pt\topsep 10pt plus 4pt minus 6pt%
 \itemsep 0pt plus 2.5pt minus 1pt}
\let\@listi\@listI
\@listi

\def\@sect#1#2#3#4#5#6[#7]#8{\ifnum #2>\c@secnumdepth%
 \def \@svsec {}\else \refstepcounter {#1}\edef \@svsec
 {\csname the#1\endcsname. \hskip .1em }\fi \@tempskipa%
 #5\relax \ifdim \@tempskipa >\z@ \begingroup #6\relax%
 \@hangfrom {\hskip #3\relax \@svsec }{\interlinepenalty%
 \@M #8.\par }\endgroup \csname #1mark\endcsname {#7}%
 \addcontentsline {toc}{#1}{\ifnum #2>\c@secnumdepth%
 \else \protect \numberline {\csname the#1\endcsname. }%
 \fi #7}\else \def \@svsechd {#6\hskip #3\@svsec #8.%
 \csname #1mark\endcsname {#7}\addcontentsline {toc}{#1}%
 {\ifnum #2>\c@secnumdepth \else \protect \numberline%
 {\csname the#1\endcsname. }\fi #7}}\fi \@xsect {#5}}

\def\section{\@startsection {section}{1}{\z@ }%
 {-3.5ex plus -1ex minus -.2ex}{2.3ex plus .2ex}{\bf }}

\def\@maketitle{%
 \newpage \null \vskip 2em
 \begin{center}{\Large\bf \@title \par }
 \vskip 1.5em
 {\large \lineskip .5em
 \begin {tabular}[t]{c}\@author
 \end{tabular}\par } \vskip .8em {{\rm June 3, 1994}}
 \end{center}\par \vskip 1.5em}

\def\thebibliography#1{%
 \section *{References.\@mkboth {REFERENCES}{REFERENCES}}%
 \list {[\arabic {enumi}]}{\settowidth \labelwidth {[#1]}%
 \leftmargin \labelwidth \advance \leftmargin \labelsep %
 \usecounter {enumi}} \def \newblock %
 {\hskip .11em plus .33em minus -.07em} \sloppy \clubpenalty 4000%
 \widowpenalty 4000 \sfcode`\.=1000\relax}

\catcode`\@=13

\newcommand{\Htriangle}[6]{
\setlength{\unitlength}{.5cm}
\begin{picture}(6,4.1)(0,-.1)
\put(0,3){\makebox(0,0){$#1$}}
\put(6,3){\makebox(0,0){$#2$}}
\put(6,0){\makebox(0,0){$#3$}}

\put(1,3){\vector(1,0){3.8}}
\put(6,2.5){\vector(0,-1){2}}
\put(1,2.5){\vector(2,-1){4}}

\put(6.9,1.5){\makebox(0,0)[r]{$#5$}}
\put(3,3.5){\makebox(0,0)[b]{$#4$}}
\put(2.3,1.05){\makebox(0,0)[l]{$#6$}}
\end{picture}
}

\newcommand{\square}[8]{
\setlength{\unitlength}{1cm}
\begin{picture}(5,3.6)
\thicklines

\put(0,3){\makebox(0,0){$#1$}}
\put(5,3){\makebox(0,0){$#2$}}
\put(0,0){\makebox(0,0){$#3$}}
\put(5,0){\makebox(0,0){$#4$}}

\put(-.5,1.5){\makebox(0,0)[r]{$#6$}}
\put(5.5,1.5){\makebox(0,0)[l]{$#7$}}
\put(2.2,0.3){\makebox(0,0)[b]{$#8$}}
\put(2.5,3.3){\makebox(0,0)[b]{$#5$}}

\put(1.7,0){\vector(1,0){1}}
\put(1,3){\vector(1,0){2.3}}
\put(0,2.5){\vector(0,-1){2}}
\put(5,2.5){\vector(0,-1){2}}
\end{picture}
}

\newcommand{\squareI}[8]{
\setlength{\unitlength}{.8cm}
\begin{picture}(5,3.6)
\thicklines

\put(0,3){\makebox(0,0){$#1$}}
\put(5,3){\makebox(0,0){$#2$}}
\put(0,0){\makebox(0,0){$#3$}}
\put(5,0){\makebox(0,0){$#4$}}

\put(-.5,1.5){\makebox(0,0)[r]{$#6$}}
\put(5.5,1.5){\makebox(0,0)[l]{$#7$}}
\put(2.5,0.5){\makebox(0,0)[b]{$#8$}}
\put(2.5,3.5){\makebox(0,0)[b]{$#5$}}

\put(1,0){\vector(1,0){2.6}}
\put(1,3){\vector(1,0){3}}
\put(0,2.5){\vector(0,-1){2}}
\put(5,2.5){\vector(0,-1){2}}
\end{picture}
}

\def\qed{\hspace*{\fill}%
\mbox{\hphantom{mm}\rule{0.25cm}{0.25cm}}\\%
\par\vskip-5mm}
\def\bk{\bf k}
\def\rada#1#2#3{{#1}_{#2},\ldots,{#1}_{#3}}

\def\LLL#1{\mbox{\rm L($#1$)}}
\def\susp{\uparrow\!}
\def\desusp{\downarrow\!}
\def\vect{\mbox{\bf Vect}}
\def\id{1\!\!1}
\def\Deltabar{\overline\Delta}
\def\sgn{\mbox{\rm sgn}}
\def\cals{{\cal S}}
\def\ext{\mbox{\large$\land$}}
\def\coland{\mbox{${}^c\!\ext$}}
\def\der{{\mbox{\rm Der}}}
\def\coder{{\mbox{\rm Coder}}}
\def\deltabar{\overline\delta}
\def\sign#1{{(-1)^{{#1}}}}
\def\lin{{\mbox{\rm Lin}}}
\def\coalg{{\mbox{\rm Coalg}}}
\def\bold#1{{\bf #1}}
\def\Cal#1{{\cal #1}}
\def\uml{{\hskip1mm\cal U}_m(L)}
\def\um#1{{\hskip1mm\cal U}_m({#1})}
\def\am{{\bf A}(m)}
\def\lm{{\bf L}(m)}
\def\ot{\otimes}
\def\fm#1{{{\cal F}_m({#1})}}
\def\S#1#2{{S_{#1,#2}}}
\def\AM{{\mbox{{\rm A(}$m${\rm )}}}}
\def\LM{{\mbox{{\rm L(}$m${\rm )}}}}
\def\End{\mbox{\rm End}}
\def\BOX{\raisebox{-.35mm}{$\Box$}}

\title{Strongly homotopy Lie algebras}
\author{Tom Lada and Martin Markl\thanks{Partially
supported by the National Research Counsel, USA}}

\begin{document}

\maketitle
\baselineskip20pt

\section{Introduction}
Strongly homotopy Lie algebras first made their appearance in a
supporting role in deformation theory~\cite{SS}.
The philosophy that every deformation
problem is directed by a differential graded Lie algebra leads, in the
context of deformation theory of a differential graded algebra $A$,
to a spectral sequence of which the $E_2$-term is naturally a strongly
homotopy Lie algebra.

For a topological space $S$, the homotopy groups $\pi_*(\Omega S)$
form
a graded Lie algebra which can be extended non-trivially (though
non-canonically) to a strongly homotopy Lie algebra which reflects
more accurately the homotopy type of $S$. The relevant operations
represent the higher order Whitehead products on $S$. In the stable
range, the basic products are given by composition and higher order
composition products; more details are given in~\cite{Smirnov}).

More recently, closed string field theory, especially in the hands
of Zwiebach and his collaborators,~\cite{Z},~\cite{WZ} has produced
a particular
strongly homotopy Lie Algebra. Lada and Stasheff~\cite{LS}
provided an exposition
of the basic ingredients of the theory of strongly homotopy Lie
algebras sufficient for the underpinnings of the physically relevant
examples. That work left open several questions naturally suggested
by comparison with the theory of differential graded Lie algebras.
The present paper addresses such questions in characteristic zero
and is complementary to what currently exists in the literature,
both physical and mathematical.

Both strongly homotopy Lie algebras and strongly homotopy associative
algebras can be expressed in terms of $n$-ary operations, respectively
$l_n$ and $\mu_n$ for all natural numbers $n\geq 1$. The defining
relations when restricted to situations with $n\leq m$ yield
corresponding
structures called $\LM$ and $\AM$ algebras respectively. Section 2 of
this work contains the basic definitions and notation regarding $\LM$
structures and is highlighted by the expected correspondence between
an $\LM$ structure on a differential graded vector space $L$, and a
degree $-1$ coderivation that is a differential on the cofree
cocommutative coalgebra generated by the suspension of $L$; this is
the content of Theorem 2.3.

In Section 3 we demonstrate the "strong homotopy" analog of the usual
relation between Lie and associative algebras. Theorem 3.1 implies
that appropriate "skew-symmetrization" of a strongly homotopy
associative algebra is a strongly homotopy Lie algebra. Here, the
condition that the field has characteristic zero is essential.
Theorem 3.3 completes the equivalence of homotopy categories: there is
a functor "universal enveloping strongly homotopy associative
algebra" from $\LM$-algebras to $\AM$-algebras which is
left adjoint to the "higher order commutators" functor.

Properties of this universal enveloping $\AM$ functor are studied
in Section 4. In particular, there exists a strict symmetric monoidal
structure
 on the category of unital $\AM$ algebras such that the universal
enveloping $\AM$ algebra functor carries a natural structure of a
unital coassociative cocommutative coalgebra with respect to this
monoidal structure. Propositions 4.1 and 4.3 contain the details
of these properties.

Section 5 is concerned with $\LM$-modules and introduces a notion
of a weak homotopy map from an $\LM$-algebra to a differential
graded Lie algebra; this generalizes certain maps considered by
Retakh~\cite{R}. A relationship between such maps and such
modules is given in Theorem 5.3.

After this paper was written, we
learned of work of Hanlon and Wachs~\cite{HW}
on $Lie_k$-algebras. Developed independently, these turn
out to be special cases of $L(k)$-algebras in which only the 'last'
map is non-zero.

We would like to express our gratitude to Jim Stasheff for his
hospitality and many fruitful conversations regarding this work.

\section{Basic definitions and notations}

All algebraic objects in the paper will be considered over a fixed
field $\bk$ of characteristic zero. We will systematically use the
Koszul sign convention meaning that whenever we commute two
``things''
of degrees $p$ and $q$, respectively,
we multiply the sign by $(-1)^{pq}$. Our conventions concerning
graded vector spaces, permutations, shuffles, etc., will follow
closely those of~\cite{MM}.

For graded indeterminates $\rada x1n$ and a permutation $\sigma\in
\cals_n$ define the {\em Koszul sign\/}
$\epsilon(\sigma)=\epsilon(\sigma;\rada x1n)$ by
\[
x_1\land\dots\land x_n = \epsilon(\sigma;x_1,\dots,x_n)
\cdot x_{\sigma(1)}\land \dots \land x_{\sigma(n)},
\]
which has to be satisfied in the free graded commutative algebra
$\ext(\rada x1n)$. Define also $\chi(\sigma) = \chi(\sigma;\rada x1n)
:= \sgn(\sigma)\cdot \epsilon(\sigma;\rada x1n)$. We say that $\sigma
\in \cals_n$ is an $(j,n-j)$-{\em unshuffle\/}, $0\leq j \leq n$, if
$\sigma(1)<\cdots<\sigma(j)$ and $\sigma(j+1)< \cdots < \sigma(n)$.

\begin{definition}
\label{shlie}
An \LLL m-structure on a graded vector space $L$ is a system
$\{l_k|\ 1\leq k\leq m,\ k<\infty\}$ of linear maps
$l_k:\otimes^k L \to L$ with $\deg(l_k) = k-2$ which are
antisymmetric in the sense that
\begin{equation}
\label{antisymmetry}
l_k(\rada x{\sigma(1)}{\sigma(k)})=\chi(\sigma)l_k(\rada x1n)
\end{equation}
for all $\sigma \in \cals_n$ and $\rada x1n\in L$, and, moreover, the
following generalized form of the Jacobi identity is supposed to be
satisfied for any $n\leq m$:
\begin{equation}
\label{Jacobi}
\sum_{i+j=n+1}\sum_\sigma
\chi(\sigma)(-1)^{i(j-1)}l_j(l_i(\rada x{\sigma(1)}{\sigma(i)}),\rada
x{\sigma(i+1)}{\sigma (n)})=0,
\end{equation}
where the summation is taken over all $(i,n-i)$-unshuffles with
$i\geq 1$.
\end{definition}

\begin{example}{\rm\
An \LLL1-algebra structure on $L$ consists of a degree $-1$
endomorphism $l_1$ and the Jacobi identity~(\ref{Jacobi}) reduces to
$l^2_1=0$, i.e.~an \LLL1-algebra is just a differential space.

An \LLL2-algebra has one more operation, a bilinear map $l_2$ which
we denote more suggestively as $[-,-]$. The antisymmetry
condition~(\ref{antisymmetry}) gives $[x,y]=-(-1)^{|x|\cdot|y|}[y,x]$
and the Jacobi condition~(\ref{Jacobi}) says that
\[
l_1([x,y])=[l_1(x),y]+(-1)^{|x|}[x,l_1(y)],
\]
in other words, an \LLL2-algebra is just an antisymmetric
nonassociative nonunital differential graded algebra.

For an \LLL3-algebra we have again one more antisymmetric operation,
$l_3$, which is the contracting homotopy for the classical Jacobi
identity:
\begin{eqnarray*}
\lefteqn{
\sign{|x|\cdot|z|}[[x,y],z]+
\sign{|y|\cdot|z|}[[z,x],y]+
\sign{|x|\cdot|y|}[[y,z],x]=}\\
&&\sign{|x|\cdot|z|+1}\!\cdot\!
\left\{
l_1l_3(x,y,z)\!+\!l_3(l_1(x),y,z)\!+\!\sign{|x|}\!\cdot\!
l_3(x,l_1(y),z)\!+\!\sign{|x|+|y|}\!\cdot \!l_3(x,y,l_1(z))
\right\}.
\end{eqnarray*}

\LLL\infty-algebras are sometimes, especially in the physical
literature, called also {\em homotopy Lie algebras\/}, but one must
beware that this name has already been reserved for $\pi_*(\Omega
S)$, the homotopy algebra of the loop space of a topological space
$S$ with the graded Lie algebra structure induced by the Samelson
product.
So, it is more appropriate to call them {\em strongly\/} homotopy
Lie algebras or {\em sh Lie\/} algebras, as in~\cite{LS}
}\end{example}

Let $L = (L,l_k)$ and $L' = (L',l'_k)$ be two \LLL m-algebras. By a
{\em map\/} of $L$ to $L'$ we mean a linear degree zero map $g: L \to
L'$ which commutes with the structure maps in the sense that
\[
g\circ l_k = l'_k \circ g^{\ot k},\ 1\leq k\leq m.
\]
Denote by $\lm$ the category of \LLL m-algebras and their
homomorphisms in the above sense. $\lm$ is an equationally given
algebraic category, a fact which we use in the next paragraph.

Let $\vect$ be the category of graded vector spaces. Denote by
$\vect^p(V,W)$ the set of linear homogeneous maps $f: V \to W$ of
degree $p$. For $V \in \vect$, let $\susp V$ (resp. $\desusp V$) be
the suspension (resp. the desuspension) of $V$, i.e. the graded
vector space defined by $(\susp V)_p = V_{p-1}$ (resp. $(\desusp V)_p
= V_{p+1}$). By $\#V$ we denote the dual of $V$, i.e. the graded
vector space $(\#V)_p :=\vect^{p}(V,{\bk})=\lin(V_{-p},{\bk})$, the
space
of linear maps from $V_p$ to $\bk$. For a graded
vector space $V$ we have the natural map $\uparrow : V \to \susp V$;
let $\uparrow^{\ot n}$
denote $\bigotimes^n \uparrow : \bigotimes^nV \to \bigotimes ^n
\susp V$, the
meaning of $\downarrow^{\ot n}$ being analogous. Notice that
$\uparrow^{\ot n}
\circ
\downarrow ^{\ot n}= \downarrow^{\ot n} \circ \uparrow^{\ot n}
= (-1)^{\frac{n(n-1)}2} \cdot \id$, as a side effect of the Koszul
sign convention.

For a graded vector space $V$, $\ext V$ will denote the free
graded commutative algebra on $V$. As usual, by $\ext^nV$ we mean the
subspace of $\ext V$ consisting of elements of length $n$, the
notations like $\ext^{\leq n}V$ having the obvious meaning. We will
need also the dual analog
of this object. Namely, for a graded vector space $W$, consider the
coalgebra $\coland W$ which, as a vector space, coincides with $\ext
W$, but the comultiplication $\Delta$ is given by $\Delta=\id\otimes
1 +\Deltabar +1\otimes\id$, where the reduced diagonal $\Deltabar$ is
defined to be
\[
\Deltabar(w_1\land\cdots\land w_n)=\sum_{1\leq j\leq n-1}\sum_\sigma
\epsilon(\sigma)(w_{\sigma(1)}\land\cdots\land w_{\sigma(j)})\otimes
(w_{\sigma(j+1)}\land\cdots\land w_{\sigma(n)}),
\]
where $\sigma$ runs through all $(j,n-j)$ unshuffles.
$\coland W$ is clearly a cocommutative (coassociative, counital)
connected coalgebra.
It has the universal property, dual to the universal property
characterizing the freeness of $\ext V$ but, as usual in the
co-algebraic world, not exactly.

To describe the universal property, introduce, for a given (counital)
coalgebra $C =(C,\Delta)$, the filtration $\{F_iC\}_{i\geq 0}$
inductively by $F_0:=0$ and $F_iC:=\{c\in C|\ \Deltabar(c)\in
F_{i-1}C\otimes F_{i-1}C\}$, $i\geq 1$. We say that $C$ is {\em
connected\/} if $C = \bigcup F_iC$. Notice that $\coland W$ itself is
connected, with $F_i\coland W = \coland^{\leq i}W$, the subspace
corresponding to $\ext^{\leq i}W$ under the identification $\coland W
= \ext W$ of graded vector spaces.

Let $\pi : \coland W \to W$ be the natural projection. The universal
property of $\coland W$ then says that, for any cocommutative
connected
coalgebra $C$ and for any linear map $\psi:C \to W$, there exists
exactly one coalgebra homomorphism $g:C \to\coland W$ such that the
diagram
\[
\Htriangle C{\coland W}Wg\pi\psi
\]
commutes.

Denote by $\iota_m:\ext^{\leq m}W\hookrightarrow \coland W$ the
obvious inclusion and, dually, let $\pi_m :\ext V \to \ext^{\leq m}V$
be the natural projection.

\begin{theorem}
\label{correspondence}
There is one-to-one correspondence between \LLL m-algebra structures
on a graded vector space $L$ and degree $-1$ coderivations $\delta$ on
the coalgebra
$\coland W$, $W := \susp L$, with the property that
$\delta^2\circ\iota_m=0$.

If the space $L$ is of finite type, then \LLL m algebra structure on
$L$ can be described also by a degree $-1$ derivation $d$ on $\ext V$,
$V=\desusp \# L$, with the property that $\pi_m\circ d^2=0$.
\end{theorem}

The first part of the theorem was, for $m=\infty$, proved
in~\cite{LS}.
The second part is, for $m=\infty$, a folk-lore result and it is
related to
the fact that the Koszul dual of the category of graded Lie algebras
is the category of graded commutative algebras~\cite{GK}. The case of
a general $m$ is an easy generalization, following the lines of the
proof of the similar statement for A($m$)-algebras,
see~\cite[Example~1.9]{MM}. We do not aim to give a proof here, but
the explicit description of the correspondences will be useful in
the sequel.

First, recall that, for a bimodule $N$ over $\ext V$, the space
$\der^p(\ext V,N)$ of degree $p$ derivations of the algebra $\ext V$
in the
bimodule $N$ has a very easy description:
\begin{equation}
\label{fuk}
\der^p(\ext V,N) \cong \vect^p(V,N).
\end{equation}
The dual statement for the coalgebra $\coland W$ and a bicomodule $M$
over $\coland W$ needs the assumption that the comodule $M$ is {\em
connected\/} meaning that, by definition, $\coland W \oplus M$ with
the obvious coalgebra structure is connected. Observe that $\coland
W$ is a connected comodule over itself. We have the following
statement.

\begin{lemma}
\label{coderivations}
\label{zebrulka}
For a connected bicomodule $N$ over the coalgebra $\coland W$ we have
an isomorphism
\[
\coder^p(M,\coland W) \cong \vect^p(M,W),
\]
induced by the correspondence $\theta \mapsto \pi\circ \theta$, $\pi :
\coland W \to W$ being the projection.
\end{lemma}

\noindent
{\bf Proof.}
Observe first that $\desusp^p M$ has a natural cobimodule structure
and
that $\coder^p(M,\coland W)\cong \coder^0(\desusp^p M,\coland W)$.
Thus we can reduce the statement of the lemma to the case $p=0$.

Let $\coalg(-,-)$ stand for the set of coalgebra maps and consider
the map
\begin{equation}
\label{opulka}
\coalg(\coland W\oplus M,\coland W)\longrightarrow
\coalg(\coland W,\coland W)
\end{equation}
given by the restriction on $\coland W$. Then $\coder^0(M,\coland W)$
obviously consists of those elements of $\coalg(\coland W\oplus
M,\coland W)$ which restrict to the identity in $\coalg(\coland
W,\coland W)$. Using the universal property of $\coland W$, the map
of~(\ref{opulka}) can be described also as
\[
\vect(\coland W\oplus M,W)\cong
\vect(\coland W,M)\oplus \vect(M,W)
\stackrel{{\mbox{\scriptsize proj.}}}{\longrightarrow}
\vect(\coland W,W)
\]
and the Lemma immediately follows.
\qed

Suppose that $\{l_k|\ 1\leq k\leq m\}$ is an \LLL m-structure on a
graded vector space $L$ as in Definition~\ref{shlie}. Let $W:=\susp
L$ and define degree $-1$ linear maps $\deltabar_k : \bigotimes^k
W\to W$ by $\deltabar_k := (-1)^{\frac{k(k-1)}2}\cdot
\mbox{$\susp \circ l_k \circ \desusp^{\otimes k}$}$,
$1\leq
k\leq
m$. Then, by the antisymmetry property~(\ref{antisymmetry}) of the
maps $l_k$, the maps $\deltabar_k$ are symmetric in the sense that
\[
\deltabar_k(\rada
x{\sigma(1)}{\sigma(k)})=\epsilon(\sigma)\deltabar_k(\rada
x1k),\ \sigma
\in \cals_k,
\]
which means that they factor to the maps (denoted by the same
symbol) $\deltabar_k: \ext^k W \to W$. By Lemma~\ref{zebrulka}
there exists exactly one coderivation $\delta \in \coder^{-1}(\coland
W)$
(= an abbreviation for $\coder^{-1}(\coland W,\coland W)$) with the
property
that
\[
\pi\circ \delta =
\left\{
\begin{array}{ll}
\deltabar_k(w),&\mbox{for}\ 1\leq k \leq m,\\
0,& \mbox{otherwise}.
\end{array}
\right.
\]
The ``Jacobi identity''~(\ref{Jacobi}) is then equivalent
to $\delta^2\circ \iota_m=0$.

On the other hand, the maps $l_k$ can be reconstructed from $\delta$
as $l_k = \desusp\circ\deltabar_k\circ\susp^{\ot k}$
with $\deltabar_k$ defined as the composition
\[
\mbox{$
\bigotimes^k W \stackrel{\mbox{\scriptsize proj.}}{\longrightarrow}
\ext^k W \hookrightarrow \coland W \stackrel\delta\longrightarrow
\coland
W \stackrel\pi\longrightarrow W.
$}
\]
This gives the correspondence of the first part of the theorem. The
description of the second one is similar. Let $V=\desusp\#L$ and
define $\overline d_k:\ext^k V \to V$ as the composition
\[
V\stackrel{\susp}{\longrightarrow}
\#L\stackrel{\#l_k}{\longrightarrow}
\mbox{$\#\bigotimes^k L$}\stackrel{\desusp^{\ot k}}{\longrightarrow}
\mbox{$\#\bigotimes^k V$}\stackrel{\mbox{\rm\scriptsize
proj.}}{\longrightarrow}\ext^kV,
\]
multiplied by $(-1)^{\frac{k(k-1)}2}$,
for $k\leq m$, and let $\overline d_k:=0$ otherwise. By~(\ref{fuk})
it defines a derivation $d\in \der^{-1}(\ext V)$ (= an abbreviation
for
$\der^{-1}(\ext V,\ext V)$). The Jacobi identity~(\ref{Jacobi}) is
then
equivalent to $d^2=0$.

On the other hand, starting from $d$, define $\overline d_k$ as the
composition
\[
V\stackrel{d}{\longrightarrow}\ext V
\stackrel{\mbox{\scriptsize proj.}}{\longrightarrow}\ext^k V
\stackrel{\mbox{\scriptsize incl.}}{\hookrightarrow}
\mbox{$\bigotimes^k V$}.
\]
Then we can reconstruct $l_k$'s as
$l_k=\#(\desusp\circ
\overline d_k\circ \susp^{\ot k})$.

\section{Symmetrization}

The usual relationship between Lie algebras and associative algebras
carries over directly to this homotopy setting.
Recall~\cite[p.~294]{St} that an
$\AM$ structure on a graded vector space $V$ is a collection
$\{ \mu_k \vert 1\le k \le m\}$ of linear maps $\mu_k:\bigotimes ^kV
\longrightarrow V$ with the degree of $\mu_k$ equal to $k-2.$ These
maps are
required to satisfy the identity
\[
\sum _{\lambda =0}^{n-1} \sum _{k=1}^{n-\lambda } (-1)^{k+\lambda
+k\lambda +nk+k(\vert a_1\vert +\dots+\vert a_{\lambda}\vert)}
\mu_{n-k+1}
(a_1,\dots,a_{\lambda},m_k(a_{\lambda +1},\dots ,a_{\lambda +k}),
a_{\lambda +k+1}, \dots,a_n)=0.
\]

We note that $\mu_1$ is a differential for $V$, $\mu_2$ is a
multiplication,
and the $\mu_k$'s are higher associating homotopies.

A homomorphism ($\AM$-map) between two $\AM$-algebras
$(V, \mu _i)$ and $(V^{\prime}, \mu_i^{\prime})$
 is a linear map $f:V\longrightarrow V^{\prime}$
of degree $0$ such that
\[
f\circ \mu_n = \mu^{\prime}_n \circ f^{\otimes n} , n=1,\dots , m.
\]
We denote by $\am$ the category of $\AM$-algebras and $\AM$-maps.
See~\cite[Example~1.9]{MM} for a thorough discussion.

We also recall that an $\AM$ structure on a graded vector
space $V$ may be described by a degree $-1$ coderivation
$\partial : {}^cTW \longrightarrow
{}^cTW$ with $\partial ^2=0$.
Here, ${}^cTW$ is the coassociative coalgebra with underlying
vector space $TW = \bigoplus_k \bigotimes^kW$ and with the reduced
diagonal $\Deltabar$ given by
\[
\Deltabar (w_1\otimes\dots \otimes w_n)=\sum_{i=1}^{n-1}
(w_1\otimes\dots\otimes w_i)\otimes (w_{i+1}\otimes\dots\otimes w_n),
\]
$W = \susp V$.
Let $\pi ^{\prime}:{}^cTW \longrightarrow W$ denote the natural
projection. The $\AM$ analog of Theorem 1.3 gives us that the $\AM$
 structure
maps $\mu_k$ can be recovered from $\partial$ by $\mu_k =
\downarrow \overline\partial_k \uparrow^{\ot k}$
where $\overline\partial _k$ is the
composition
\[
\mbox{$\bigotimes$}^kW
\stackrel{\mbox{\scriptsize incl.}}{\hookrightarrow} {}^cTW
\stackrel{\partial}{\longrightarrow} {}^cTW \stackrel {\pi ^{\prime}}
{\longrightarrow} W.
\]
Details may be found in~\cite{GL}.

\begin{theorem}
An $\AM$-structure $\lbrace \mu_n:\bigotimes ^nV\longrightarrow
V\rbrace$
on the graded vector space $V$ induces an $\LM$-structure
$\lbrace l_n:\bigotimes ^nV \longrightarrow V \rbrace$ where
\[
l_n(v_1\otimes \dots \otimes v_n):=\sum_{\sigma\in \cals_n}\chi
(\sigma )
\mu_n(v_{\sigma (1)}\otimes \dots \otimes v_{\sigma (n)}),\ 1\leq
n\leq m.
\]
This correspondence defines a functor $(-)_L:\am \longrightarrow \bold
L(m)$.

\end{theorem}

\noindent
{\bf Proof.} Consider the injective coalgebra map $S :{}^c\!\bigwedge
W
\longrightarrow {}^cTW$ given by $$S(w_1 \wedge \dots \wedge w_n)=
\sum _{\sigma \in S_n}\epsilon (\sigma )(w_{\sigma (1)}\otimes \dots
\otimes w_{\sigma (n)}).$$
Using Lemma 1.4, we extend the linear map $\pi ^{\prime}\partial S:
{}^c\!\bigwedge W
\longrightarrow W$ to the unique coderivation $\delta :{}^c\!\bigwedge
W
\longrightarrow {}^c\!\bigwedge W$ which has the property that
$\pi \delta = \pi ^{\prime} \partial S$. Since $\delta ^2 :
{}^c\!\bigwedge W \longrightarrow {}^c\!\bigwedge W$ is a
coderivation, to show
that
$\delta ^2=0$ we need only show that $\pi \delta ^2 =0$. But, $\pi
\delta ^2 = \pi ^{\prime}\partial S \delta$ which is equal to $0$
if $S \delta = \partial S$. Since S is a coalgebra map, $S \delta$
and $\partial S \in $ Coder$({}^c\!\bigwedge W, {}^cTW)$ and so we
need examine
only $\pi ^{\prime}S\delta$ and $\pi ^{\prime}\partial S$; since
$\pi ^{\prime}S = \pi, \pi ^{\prime}S \delta = \pi \delta$ whereas
$\pi ^{\prime} \partial S = \pi \delta$ by definition. The resulting
$\LM$-structure on $V$ now follows from Theorem 1.3.
In addition, if $f:(V,\mu_n) \longrightarrow
(V^{\prime},\mu_n^{\prime})$ is an
$\AM$-map so that $\mu^{\prime}_n \circ f^{\otimes n} = f \circ
\mu_n$,
then $f \circ l_n(v_1\otimes\dots\otimes v_n) = \sum
_{\sigma\in\cals_n}
\chi (\sigma ) f \circ \mu_n(v_{\sigma (1)}\otimes\dots\otimes
v_{\sigma (n)})
= \sum_{\sigma\in\cals_n } \chi (\sigma ) \mu^{\prime}_n \circ
f^{\otimes
n}(v_{\sigma (1)}\otimes\dots\otimes v_{\sigma (n)})
= l^{\prime}_n \circ f^{\otimes n}(v_1\otimes\dots\otimes v_n)$
which shows the functoriality of our construction.
\qed

\vskip-4mm
\begin{remark}{\rm\
It should be clear that for $n=2$,
$l_2(v_1\otimes v_2)=\mu_2(v_1\otimes v_2) - (-1)^
{\vert v_1 \vert \vert v_2 \vert}
\mu_2 (v_2 \otimes v_1)$
is the usual graded commutator. For $n>2$, the $l_n$'s are the
appropriate symmetrization of the associating homotopies.

If $A = (A,\partial,\cdot)$ is an associative differential graded
algebra considered in an obvious way as an $\AM$-algebra
for some $m\geq 2$
(see~\cite[Example~1.5]{MM}), then $A_L$ is the usual commutator Lie
algebra associated to $A$.
}\end{remark}

The following proposition follows from the fact that
$(-)_L:\am \to \lm$
is an algebraic functor and thus has a left adjoint.

\begin{theorem}
There is a functor $\Cal U_m : \bold L(m)
\longrightarrow \am$ that is left adjoint to $(-)_L$.
$\Cal U_m$ is called the universal enveloping $\AM$-algebra functor
for $\LM$-algebras.
\end{theorem}

There exists another construction of a ``universal enveloping
algebra'' which gives, for any sh~Lie algebra $L$, an {\em
associative\/} ({\em not\/} A($\infty$)) algebra characterized by
a certain universal property with respect to $L$-modules,
see~\cite{HS}.
We used the name ``universal enveloping $\AM$-algebra'' instead of
just ``universal enveloping algebra'' to distinguish between these two
constructions.

There is a description of $\Cal U_m(L)$ that is analogous to the
classical
description of the universal enveloping algebra of a Lie algebra.
We begin with a
graded vector space $L$ with its $\LM$-structure $\lbrace l_n
\rbrace$. Let $\Cal F_m(L)$ be the free $\AM$-algebra
generated by the vector space $L$ with $\AM$-structure maps
denoted by $\lbrace \mu _n\rbrace$. Let $I$ denote the ideal in
$\Cal F_m(L)$ generated by the relations
\[
\sum_{\sigma \in S_n} \chi (\sigma) \mu_n (\xi _{\sigma (1)},\dots ,
\xi _{\sigma(n)})
= l_n(\xi _1, \dots , \xi_n)
\]
where $\xi_1,\dots,\xi_n\in L$.
Let $\Cal U_m(L) = \Cal F_m(L) / I$ and $j:L \longrightarrow \Cal
U_m(L)$
be the natural inclusion. $\Cal U_m(L) $ is then universal in the
following sense: given a linear map $f:L \longrightarrow A$ where $A$
is
an $\AM$-algebra such that $f : L \longrightarrow A_L$ is
an $\LM$-homomorphism, there is a unique $\AM$-map
$\hat f:\Cal U_m(L) \longrightarrow A$ such that $\hat f\circ j = f$.
To see this, note that there is a unique homomorphism of
$\AM$-algebras, $\hat f :\Cal F_m(L) \longrightarrow A$ such that
$\hat f
\circ j = f$ since $\Cal F_m(L)$ is free. We need only check that
$\hat f(I) = 0$. We denote the
$\AM$-structure on $A$ by $\lbrace \hat \mu_n\rbrace$ and its
corresponding commutator $\LM$-structure by $\lbrace
\hat l_n\rbrace$.

We apply $\hat f$ to each side of the equation that defines the
ideal $I$ and obtain
\begin{eqnarray*}
\lefteqn{
\hat f (\sum _{\sigma} \chi (\sigma )\mu _n(\xi _{\sigma (1)},\dots,
\xi _{\sigma (n)})) = \sum _{\sigma} \chi (\sigma)\hat f \mu _n
(\xi _{\sigma (1)}, \dots \xi _{\sigma (n)})=}
\\
&&= \sum _{\sigma} \chi (\sigma)
\mu_n \hat f(\xi _{\sigma (1)}, \dots,
\xi _{\sigma (n)}) = \sum _{\sigma} \chi (\sigma ) \mu_n(f\xi
_{\sigma (1)},
\dots , f\xi _{\sigma (n)})
\end{eqnarray*}
on the left. On the other side we have
\[
\hat f (l_n(\xi _1, \dots , \xi _n)) = f(l_n(\xi _1, \dots , \xi
_n))=\hat l_n(f\xi _1, \dots ,f\xi _n) = \sum_{\sigma} \chi (\sigma )
\mu_n(f\xi _{\sigma (1)},\dots , f\xi _{\sigma (n)}).
\]
This shows that $\hat f(I) =0$, therefore $\hat f$ factors to the
requisite map $\Cal U_m(L) \to A$.

\section{Some properties of $\uml$}

The aim of this section is to show the existence of a strict
symmetric monoidal structure on the category $\am$ of unital
\AM-algebras such that the
universal enveloping $\AM$-algebra constructed in the previous
section carries a natural structure of a unital coassociative
cocommutative coalgebra with respect to this
monoidal structure, c.f.~the classical analog of this
result~\cite{HS}.

Let $A$ and $B$ be two A($m$)-algebras. Choose free presentations $A
= \fm {X_A}/(R_A)$ and $B=\fm {X_B}/(R_B)$. Then define
\begin{equation}
\label{box}
A\BOX B := \fm{X_A\oplus X_B}/(R_A,\S AB,R_B),
\end{equation}
where $\S AB$ is the ideal generated by the relations
\[
\sum_{\sigma \in
S_n}\chi(\sigma;x_1,\ldots,x_n)\cdot\mu_n(x_{\sigma(1)},\ldots,x_{%
\sigma(n)})=0,
\]
with $x_{i_1},\ldots,x_{i_s} \in X_A$, $x_{j_1},\ldots,x_{j_t} \in
X_B$, where
$\{i_1,\ldots,i_s,j_1,\ldots,j_t\}$ is a decomposition of
$\{1,\ldots,n\}$, $s,t\geq 1$, and $s+t=n$.

In the following proposition, {\bf 1} will denote the trivial unital
\AM-algebra, ${\bf 1}=({\bf k},\mu_i)$, with
$\mu_i(1\otimes\cdots\otimes1)=1$ for $i=2$ and
$\mu_i(1\otimes\cdots\otimes1)=0$ otherwise.

\begin{proposition}
\label{symmetry}
The operation $\BOX$ introduced above induces on the category $\am$
the structure of a strict symmetric monoidal category with {\bf 1} as
the unit object.
\end{proposition}

\noindent
{\bf Proof.}
Notice that the formula~(\ref{box}) gives a well-defined functor
$\BOX :\am \times\am \to \am$. The obvious fact that $\S AB = \S BA$
gives the symmetry $s : A\BOX B \to B \BOX A$.

Let $C = \fm{X_C}/(R_C)$ be a third A($m$)-algebra. We have, by
definition,
\begin{eqnarray*}
A\BOX(B\BOX C) &=& \fm{X_A\oplus X_B\oplus X_C}/(R_A,R_B,R_C,\S
A{B\oplus C},\S BC),
\\
(A\BOX B) \BOX C &=& \fm{X_A\oplus X_B\oplus X_C}/(R_A,R_B,R_C,\S
AB,\S {A\oplus B}C).
\end{eqnarray*}
On the other hand, clearly
\[
(\S A{B\oplus C},\S BC)=(\S AB,\S {A\oplus B}C)=(\S AB,\S BC,\S CA)
\]
which easily gives the ``associativity isomorphism''
$\alpha_{A,B,C}: A\BOX (B\BOX C) \to (A\BOX B)\BOX C)$.

Finally, if {\bf 0} denotes the trivial vector space, then ${\bf 1}=
\fm{{\bf 0}}/({\bf 0})$ and we see immediately that $A\BOX {\bf 1}=
{\bf 1}\BOX A = A$. The reader may easily verify that the structures
constructed above satisfy the axioms of a strict symmetric monoidal
category as they are listed, for example, in~\cite{McL}.
\qed

Let $L'= (L',l'_i)$ and $L''=(L'',l''_j)$ be two L($m$)-algebras.
Define their {\em direct product\/} $L'\times L''= (L'\oplus
L'',l_n)$ by
\[
l_n(\xi_1,\ldots,\xi_n):=
\left\{
\begin{array}{lll}
l_n'(\xi_1,\ldots,\xi_n),& \mbox{if all $\xi_i\in L'$,}
\\
l_n''(\xi_1,\ldots,\xi_n),& \mbox{if all $\xi_i\in L''$, and}
\\
0,& \mbox{otherwise}.
\end{array}
\right.
\]

\begin{proposition}
For any two \LM-algebras $L'$ and $L''$ there is a natural
isomorphism
\[
\um{L'\oplus L''} \cong \um{L'}\BOX\um{L''}
\]
of \AM-algebras.
\end{proposition}

{\bf Proof.}
Using the description of the universal enveloping A($m$)-algebra as
it is given at the end of the previous paragraph, we have
\[
\um{L'\times L''} = \fm{L'\oplus L''}/(I),
\]
with I is the ideal generated by the relations
\begin{equation}
\label{boo}
\sum_{\sigma \in S_n} \chi (\sigma)\cdot \mu_n (\xi_{\sigma(1)},\dots
,
\xi_{\sigma(n)})
= l_n(\xi _1, \dots , \xi_n),
\end{equation}
where $\xi_1,\cdots,\xi_n \in L'\oplus L''$. It is immediate to see
that $I = (R',S,R'')$, where $R'$ (resp.~$R''$) is generated by the
relations~(\ref{boo}) with $\xi_i\in L'$ (resp.~$\xi_i\in L''$) and
$S=\S{L'}{L''}$. The proposition now follows from the definition of
the $\BOX$-product.
\qed

Let $L$ be an L($m$)-algebra and let $\delta : L \to L\times L$
be the homomorphism given by $\delta(\xi):= \xi \oplus\xi$. This map
induces, by the functoriality of $\um-$, the $\am$-map $\Delta :\uml
\to \um{L\times L} = \uml\BOX\uml$.

\begin{proposition}
The homomorphism $\Delta :\uml \to \uml\BOX\uml$ induces on $\uml$
the structure of a cocommutative coassociative coalgebra in the
monoidal category $(\am,\BOX,{\bf 1})$, the counit given by the
augmentation $\epsilon : \uml\to {\bf 1}$.
\end{proposition}

\noindent
{\bf Proof.}
The coassociativity of $\Delta$ means that the diagram
\[
\square%
{\uml}{\uml\BOX\uml}{\uml\BOX\uml}{\uml\BOX\uml\BOX\uml}%
{\Delta}{\Delta}{\Delta\BOX\id}{\id\BOX\Delta}
\]

\vskip2mm\noindent
commutes. But this diagram is obtained by applying the functor $\um-$
on the diagram

\vskip2mm
\[
\setlength{\unitlength}{.8cm}
\squareI%
{L}{L\times L}{L\times L}{L\times L\times L}%
{\delta}{\delta}{\delta\times\id}{\id\times\delta}
\]
which obviously commutes.

The fact that $\epsilon$ is a left counit means, by definition, that
the composition
\[
\uml \stackrel{\Delta}{\longrightarrow} \uml\BOX\uml
\stackrel{\epsilon\BOX\id}{\longrightarrow}{\bf 1}\BOX\uml
\stackrel{\cong}{\longrightarrow} \uml
\]
is the identity map. But this composition is obtained by applying the
functor $\um-$ on the composition
\[
L \stackrel{\delta}{\longrightarrow}L\times L
\stackrel{0\times \id}{\longrightarrow}{\bf 0}\times L
\stackrel{\cong}{\longrightarrow} L
\]
which is plainly the identity. The proof that $\epsilon$ is also a
right
counit is the same.
\qed

\vskip-4mm
\begin{remark}
{\rm\
The arguments of this section seem to suggest a general scheme valid
for
all universal-enveloping-algebra-like functors ${\cal U}: {\cal A}\to
{\cal B}$ (we know three examples of such functors, the
``classical'' universal enveloping algebra functor for Lie
algebras~\cite{HS},
the universal enveloping algebra functor for Leibniz algebras
of~\cite{L}
and, of course, our functor $\um-$).

To construct a coassociative coalgebra structure on ${\cal U}(L)$,
look for a strict monoidal structure $-\odot-$ on the category ${\cal
B}$ having the property that the functors ${\cal U}(-\times-)$ and
${\cal U}(-)\odot {\cal U}(-)$ ($-\times-$ denoting the direct
product in the category ${\cal A}$) are naturally equivalent.
Then the
coassociative comultiplication on ${\cal U}(L)$ is induced by the
diagonal map $\delta:L\times L\to L$.
Examples of these suitable monoidal structures are: the tensor
product $-\otimes-$ for the classical universal enveloping algebra
functor, the free product for the universal enveloping algebra
functor for Leibniz algebras and, of course, the operation $-\BOX-$
for our functor
$\um-$.
}
\end{remark}

\section{$\LM$-modules}

We introduce two concepts in this section. The first is that of
left modules over $\LM$-algebras. The second idea involves
homomorphisms from $\LM$-algebras to differential graded Lie
algebras. It is not surprising that these two ideas are closely
related. We begin with the following

\begin{definition}
\label{snuff}
Let $L=(L,l_i)$, be an $\LM$-algebra, and let $M$ be a differential
graded vector space with differential denoted by $k_1$. Then a left
$L$-module structure on $M$ is a collection $\{k_n| 1\leq n\leq m,\
n< \infty\}$ of linear maps of degree $n-2$,
$$
k_n:\bigotimes ^{n-1}L\otimes M \longrightarrow M,
$$
such that
\begin{equation}
\label{module}
\sum_{i+j=n+1}\sum_{\sigma}\chi (\sigma ) (-1)^{i(j-1)}k_j(k_i(
\xi_{\sigma(1)},\dots ,\xi_{\sigma(i)}),\xi_{\sigma(i+1)},\dots,
\xi_{\sigma(n)})=0
\end{equation}
where $\sigma$ ranges over all $(i,n-i)$ unshuffles, $\xi_1,\dots,
\xi_{n-1}\in L$ and $\xi_n \in M$.
\end{definition}

Several comments are in order. We assume that $\xi_n \in M$ while
the other $\xi_i$'s $\in L$, and then according to the definition
of $(i,n-i)$ unshuffles, it follows that either $\xi_{\sigma(i)}=
\xi_n$ or $\xi_{\sigma(n)}=\xi_n$. In the first case then, we
define
$$
k_j(k_i(\xi_{\sigma(1)},\dots,
\xi_{\sigma(i)}),\xi_{\sigma(i+1)},\dots,
\xi_{\sigma(n)})
:=\alpha\cdot k_j(\xi_{\sigma(i+1)}, \dots, \xi_{\sigma(n)},
k_i(\xi_{\sigma(1)}, \dots, \xi_{\sigma(i)}))
$$
where
$$
\alpha = (-1)^{j-1}\cdot (-1)^{
(i+\sum_{k=1}^i |\xi_{\sigma(k)}|)\cdot
(\sum^n_{k=i+1}\vert \xi_{\sigma(k)}\vert)}
$$
according to the Koszul sign convention. In the second case,
i.e. when $\xi_{\sigma(i)}\in L$, we take $k_i=l_i$.

It is not difficult to show that $L$-modules in the above sense are
abelian group objects in the slice category $\lm/L$ of
\LLL m-algebras over $L$.

Of course, the fundamental example of such a structure occurs in
the situation when $M=L$ and each $k_i=l_i$, i.e., $L$ is an
$\LM$-module over itself. Definition~\ref{snuff} should be compared
with the definition of a module (resp.~balanced module) over an
A($m$)-algebra (resp.~balanced A($m$)-algebra) as it was given
in~\cite[1.10]{MM}.

We next consider maps from \LLL m-algebras to differential
graded Lie algebras.

\begin{definition}
\label{shuff}
Let $L= (L,l_i)$ be an $\LM$-algebra and
$A= (A, \partial _A, [-,-])$ a differential graded Lie algebra. A
weak $\LM$-map from $L$ to $A$ is a collection
$\{f_n|\ 1\leq n\leq m-1,\ n< \infty\}$ of skew symmetric
linear maps $f_n:\bigotimes^nL \longrightarrow A$ of degree
$n-1$ such that
\begin{eqnarray}
\label{homomorphism}
\lefteqn{
\hphantom{mm}\partial _A f_n(\xi_1,\dots,\xi_n)
\!+\hskip-5mm\sum_{j+k=n+1}\sum_{\sigma}\chi(\sigma)(-1)^{k(j-1)+1}
f_j(l_k(\xi_{\sigma(1)},
\dots,\xi_{\sigma(k)}),\xi_{\sigma(k+1)},\dots,\xi_{\sigma(n)})}
\\
\nonumber
&&+\hskip-2mm\sum_{s+t=n}\sum_{\tau}\chi(\tau)(-1)^{s-1}\cdot
(-1)^{(t-1)(\sum_{p=1}^s\vert \xi_{\tau
(p)}\vert)}\cdot[f_s(\xi_{\tau(1)},\dots,\xi_{\tau(s)}),
f_t(\xi_{\tau(s+1)},\dots,\xi_{\tau(n)})]=0
\end{eqnarray}
where $\sigma$ runs through all $(k,n-k)$ unshuffles and $\tau$
runs through all $(s,n-s)$ unshuffles such that $\tau(1)<\tau(s+1)$,
 and $[-,-]$ denotes the
graded bracket on $A$, $\xi_1, \dots, \xi_n\in L$.
\end{definition}

\begin{remark}{\rm\
Let $L=(L,l_i)$
and $L'=(L',l'_i)$ be two \LLL \infty-algebras.
Let $(\coland W,\delta)$ and
$(\coland W',\delta')$ be the corresponding differential graded
coalgebras as in Theorem~\ref{correspondence}. We may say that a
{\em weak map\/} from $L$ to $L'$ is a differential graded coalgebra
homomorphism $\psi : (\coland W,\delta) \to (\coland W',\delta')$. We
may also say that such a weak map is a (strict)
{\em map\/} from $L$ to $L'$
if $\psi (\coland ^n W)\subset \coland ^n W'$ for each $n\geq 1$. It
is almost obvious to see that this definition is equivalent to the
definition of a map as it was given in Section 2. Definition~\ref{shuff}
is then equivalent in the special case $L'=A$,
$l'_1= \partial_A$, $l'_2 = [-,-]$, and $l'_k=0$ for $k\geq 3$,
to the definition of a weak map above. The case of a general
$m<\infty$ can be discussed in a similar way.

In the special case of $L$ having a strict differential graded
Lie structure, our
definition agrees with the definition of an $(m-1)$-homotopically
multiplicative map studied by Retakh in~\cite{R}.
}\end{remark}

The following theorem shows that we have the usual
relationship between
homomorphisms and module structures. Let $\End(M)$ denote the
graded associative algebra of linear maps from $M$ to $M$ with
product given by composition and differential induced by the
differential $k_1$ on $M$. Let us denote by $\End(M)_L$ the
differential graded Lie algebra associated to the differential graded
associative algebra $\End(M)$.

\begin{theorem}
\label{corresp}
Suppose that $L=(L,l_i)$ is an $\LM$-algebra and that $M =(M,k_1)$ is
a
differential graded vector space. Then there exists a natural
one-to-one
correspondence between $L$-module
structures on $M$ and weak \LLL m-maps $L\to \End(M)_L$.
\end{theorem}

\noindent
{\bf Proof.}
For each module structure map $k_n:\bigotimes^{n-1}L\otimes M
\to M$ we define $f_{n-1}:\bigotimes^{n-1}L\to \End(M)$ by
$$
f_{n-1}(\xi_1,\dots,\xi_{n-1})(m):=(-1)^{n+1}\cdot
k_n(\xi_1,\dots,\xi_{n-1},m).
$$
Let us consider the defining equation~(\ref{module}) multiplied by
$(-1)^{n+1}$:
$$
\sum_{i+j=n+1}\sum_{\sigma}\chi(\sigma)(-1)^{j(i+1)}\cdot
k_j(k_i(\xi_{\sigma(1)},
\dots,\xi_{\sigma(i)}),\xi_{\sigma(i+1)},\dots,\xi_{\sigma(n)})=0.
$$
We may obviously split the summation into four parts, the first one
with $\sigma(n)=n$ and $j>1$, the second one with $\sigma(i)=n$ and
$i,j > 1$, the third one with $\sigma(i)=n$, $i=1$ and $j> 1$, and
the fourth one with $j=1$. We obtain
\begin{eqnarray*}
0&=&
\sum_{\mbox{\scriptsize$
\begin{array}{c}
i\!+\!j\!=\!n\!+\!1
\\
j\!>\!1
\end{array}$}}
\sum_{\sigma(n)=n}\chi(\sigma)(-1)^{j(i+1)}\cdot
k_j(l_i(\xi_{\sigma(1)},\dots,\xi_{\sigma(i)}),
\xi_{\sigma(i+1)},\dots,\xi_n)
\\
&+&\sum_{\mbox{\scriptsize$
\begin{array}{c}
i\!+\!j\!=\!n\!+\!1
\\
\scriptsize j\!>\!1, i\!>\!1
\end{array}$}}
\sum_{\sigma(n)\neq n}\chi(\sigma)(-1)^{j(i+1)}\cdot
k_j(k_i(\xi_{\sigma(1)},\dots,\xi_{\sigma(i)}),\xi_{\sigma(i+1)}
,\dots,\xi_{\sigma(n)})
\\
&-&(-1)^{n} (-1)^{|\xi_n|(\sum_{p=1}^{n-1}|\xi_\sigma(p)|)}
\cdot k_n(k_1(\xi_n),\xi_1,\ldots,\xi_{n-1})
+(-1)^{n+1}\cdot k_1(k_n(\xi_1,\dots,\xi_n)).
\end{eqnarray*}
The first term of the right-hand side may then be written as
$$
\sum_{i+j=n+1}\sum_{\sigma}\chi(\sigma)(-1)^{ij+1}\cdot f_{j-1}
(l_i(\xi_{\sigma(1)},\dots,\xi_{\sigma(i)}),\xi_{\sigma(i+1)},
\dots,\xi_{\sigma(n-1)})(\xi_n)
$$
which is easily seen to correspond to the first part
of~(\ref{homomorphism})
(after the substitution $n\mapsto n+1$, $j\mapsto j+1$ and
$i\mapsto k$).

The second term of the relation requires a more subtle examination.
For a fixed $(i,n-i)$ unshuffle $\sigma$ with $\sigma(i)=n$,
we have the corresponding $(n-i+1,i-1)$ unshuffle
$\sigma ^{\prime}$ with $\sigma^{\prime}
(n-i+1)=n$ given by $\sigma':
(1,\ldots,n)\mapsto (\sigma(i+1),\ldots,\sigma(n),
\sigma(i),\sigma(1),\ldots,\sigma(i-1))$.
We then pair the terms that are indexed by these two
unshuffles and reindex the sum with just one of the unshuffles, say
$\sigma$, where $\sigma$ is chosen so that
$\sigma(1)<\sigma(i+1)$, to obtain
\begin{eqnarray*}
\lefteqn{
\chi(\sigma)(-1)^{j(i+1)}\cdot k_j(k_i(\xi_{\sigma(1)},\dots,
\xi_{\sigma(i)}),\xi_{\sigma(i+1)},\dots,\xi_{\sigma(n)})}
\\
&&+\chi(\sigma)(-1)^{i(j+1)}\beta \cdot
k_i(k_j(\xi_{\sigma(i+1)},\dots,
\xi_{\sigma(n)},\xi_{\sigma(i)}),\xi_{\sigma(1)},\dots,\xi_{%
\sigma(i-1)})
\end{eqnarray*}
where
\[
\beta =(-1)^{ij-1}\cdot
(-1)^{\vert \xi_{\sigma(i)}\vert\cdot( \sum^n_{q=i+1}\vert
\xi_{\sigma(q)}\vert)+(\sum^{i-1}_{p=1}\vert\xi_{\sigma(p)}\vert)
(\sum^n_{q=i}\vert\xi_{\sigma(q)}\vert)}
\]
defined by $\chi(\sigma') =\beta \cdot \chi(\sigma)$
is the sign adjustment that allows us to relate the
two unshuffles to the same permutation $\sigma$.

We rewrite this sum as
\begin{eqnarray*}
\lefteqn{
\chi(\sigma)(-1)^{j(i+1)}\cdot \alpha_1
\cdot k_j(\xi_{\sigma(i+1)},\dots,
\xi_{\sigma(n)},k_i(\xi_{\sigma(1)},\dots,\xi_{\sigma(i)}))}
\\
&&+\chi(\sigma)(-1)^{i(j+1)}\cdot
\beta\alpha_2\cdot k_i(\xi_{\sigma(1)},\dots,
\xi_{\sigma(i-1)},k_j(\xi_{\sigma(i+1)},\dots,\xi_{\sigma(n)},
\xi_{\sigma(i)}))
\end{eqnarray*}
where
$$
\alpha_1=(-1)^{j-1}\cdot(-1)^{(i+\sum_{p=1}^i |\xi_{\sigma(p)}|)\cdot
(\sum^n_{q=i+1}\vert \xi_{\sigma(q)}\vert)}
$$
is defined by $\chi(k_i(\xi_{\sigma(1)},\dots,
\xi_{\sigma(i)}),\xi_{\sigma(i+1)},\dots,\xi_{\sigma(n)}) =
\alpha_1\cdot\chi(\xi_{\sigma(i+1)},\dots,
\xi_{\sigma(n)},k_i(\xi_{\sigma(1)},\dots,\xi_{\sigma(i)}))$ and
$$
\alpha_2=(-1)^{i-1}\cdot(-1)^{(j +\sum_{q=i}^n |\xi_{\sigma(q)}|
)\cdot (\sum^{i-1}_{p=1}\vert \xi_{\sigma(p)}\vert )}
$$
is defined by
\begin{eqnarray*}
\lefteqn{
\chi(k_j(\xi_{\sigma(i+1)},\dots,
\xi_{\sigma(n)},\xi_{\sigma(i)}),\xi_{\sigma(1)},\dots,\xi_{%
\sigma(i-1)})=\hphantom{mmmmm}}
\\
&&\hphantom{mmmmm}= \alpha_2\cdot\chi(\xi_{\sigma(1)},\dots,
\xi_{\sigma(i-1)},k_j(\xi_{\sigma(i+1)},\dots,\xi_{\sigma(n)},
\xi_{\sigma(i)})).
\end{eqnarray*}
When the correspondence $k_n \leftrightarrow (-1)^{n+1}\cdot f_{n-1}$
is made explicit in the above, we arrive at
\begin{eqnarray*}
\lefteqn{
\chi(\sigma)\cdot\varphi
\cdot\left\{ (f_{i-1}(\xi_{\sigma(1)},\ldots,\xi_{\sigma(i-1)})
\circ f_{j-1}(\xi_{\sigma(i+1)},\ldots,
\xi_{\sigma(n)})\right.\hskip5mm}
\\
&&\hskip5mm\left.\hskip2mm-\gamma f_{j-1}
(\xi_{\sigma(i+1)},
\ldots,\xi_{\sigma(n)}) \circ
f_{i-1}(\xi_{\sigma(1)},\ldots,\xi_{\sigma(i-1)}))(\xi_{\sigma(i)})
\right\}
\end{eqnarray*}
where $\varphi:= (-1)^{i+j} \cdot (-1)^
{\vert \xi_{\sigma(i)}\vert
(\sum^n_{q=i+1}
\vert \xi_{\sigma(q)}\vert)}\cdot
(-1)^{j(\sum^{i-1}_{p=1}\vert \xi_{\sigma(p)}\vert)}$ and
\[
\gamma=(-1)^{\vert f_{i-1}(\xi_{\sigma(1)},\ldots,
\xi_{\sigma(i-1)})\vert\cdot \vert f_{j-1}(\xi_{\sigma(i+1)},
\ldots,\xi_{\sigma(n)})\vert}=
(-1)^{(i+\sum_{p=1}^{i-1}|\xi_{\sigma(p)}|)\cdot
(j+\sum_{q=i+1}^{n}|\xi_{\sigma(q)}|)}
\]
is the sign required for the commutator. Let us define an
$(i-1,n-i-1)$-unshuffle $\tau$ by
\[
\tau(k):=
\left\{
\begin{array}{ll}
\sigma(k),&\mbox{ for $1\leq k\leq i-1$, and}
\\
\sigma(k+1),&\mbox{ for $i\leq k\leq n-1$}.
\end{array}
\right.
\]
Then $\chi(\sigma)= (-1)^j\cdot (-1)^{\vert \xi_{\sigma(i)}
\vert\cdot (\sum^n_{q=i+1}\vert \xi_{\sigma(q)}\vert)}
\cdot \chi(\tau)$ and the substitution $\sigma \mapsto \tau$ enables
us to write the above expression as
\[
\chi(\tau)\cdot(-1)^{j(\sum^{i-1}_{p=1}\vert
\xi_{\tau(p)}\vert)} \cdot(-1)^i\cdot
[f_{i-1}(\xi_{\tau(1)},\ldots,\xi_{\tau(i-1)}),
f_{j-1}(\xi_{\tau(i)},\ldots,\xi_{\tau(n-1)})](\xi_{n})
\]
which corresponds, after the substitution $i\mapsto s+1$,
$j\mapsto t+1$ and $n\mapsto n+1$,
to the third term of~(\ref{homomorphism}).

The remaining two terms can be written as
$$
k_1 \circ f_{n-1}(\xi_1,\ldots,\xi_{n-1})(\xi_n)-
(-1)^{(n+\sum_{k=1}^{n-1}|\xi_k|)}\cdot
f_{n-1}(\xi_1,\ldots,\xi_{n-1})(k_1(\xi_n))
$$
which is the differential in $\End(M)_L$ applied to
$f_{n-1}(\xi_1,\ldots,\xi_{n-1})$,
i.e.~the first term of~(\ref{homomorphism}) (after the substitution
$n\mapsto n+1$).
\qed

We note that the pairing of the unshuffles in the above proof
leads to the same index set called "regular sequences" in~\cite{R}.

We believe that an analog of Theorem~\ref{corresp} holds also for
modules (resp.~balanced modules) over an
A($m$)-algebra (resp.~balanced A($m$)-algebra)~\cite[1.10]{MM}.

\catcode`\@=11
\noindent
T.~L.: Math.~Department, NCSU, Raleigh, NC 27695 - 8205, USA
\hfill\break\noindent
\hphantom{T.~L.:} email: {\bf lada@math.ncsu.edu}

\noindent
M.~M.: Mathematical Institute of the Academy, \v Zitn\'a 25, 115 67
Praha 1, Czech Republic,\hfill\break\noindent
\hphantom{M.~M.:} email: {\bf markl@earn.cvut.cz}
\catcode`\@=13


\frenchspacing
\begin{thebibliography}{10}

\bibitem{GK}
{V. Ginzburg and M. Kapranov.}
\newblock Koszul duality for operads.
\newblock Preprint.

\bibitem{GL}
{V.K.A.M. Gugenheim and L. Lambe.}
\newblock Perturbation theory in differential homological algebra~I.
\newblock Ill. J. Math. 33 no.~4 (1989), 566-582.

\bibitem{HS}
P.J. Hilton and U. Stammbach.
\newblock A Course in Homological Algebra.
\newblock Graduate texts in Mathematics~4,
Springer-Verlag, 1971.

\bibitem{HSch}
V. Hinich and V. Schechtman.
\newblock Homotopy Lie algebras. Preprint.

\bibitem{HW}
P. Hanlon and M. Wachs.
\newblock $Lie_k$-algebras. Preprint.

\bibitem{LS}
{T. Lada and J. Stasheff.}
\newblock Introduction to sh Lie algebras for physicists.
\newblock International Journal of Theoretical Physics Vol. 32,
No. 7 (1993), 1087--1103.

\bibitem{L}
{J. Loday.}
\newblock Une version non commutative des alg\'ebres de Lie:
les alg\'ebres de Leibniz, Enseign. Math., to appear.


\bibitem{MM}
{M. Markl.}
\newblock A cohomology theory for {A($m$)-algebras} and
applications.
\newblock Journ. Pure Appl. Algebra 83 (1992),
141--175.

\bibitem{McL}
S. MacLane.
\newblock Categories for the working mathematician.
Springer-Verlag 1971.

\bibitem{R}
{V.S. Retakh.}
\newblock Lie-Massey brackets and $n$-homotopically multiplicative
maps of differential graded Lie algebras.
\newblock Journ. Pure Appl. Algebra 89 (1993), 217--229.

\bibitem{SS}
M. Schlessinger and J. D. Stasheff.
\newblock Deformation theory and rational homotopy type.
\newblock Publ. Math IHES (to appear).

\bibitem{Smirnov}
V.A. Smirnov.
\newblock A general algebraic approach to the problem of describing
the second term of the Adams spectral sequence and stable homotopy
groups.
\newblock Preprint.

\bibitem{St}
J.D. Stasheff.
\newblock Homotopy associativity of {H-spaces} {II.}
\newblock {\em Trans. Amer. Math. Soc.}, {\bf 108}(1963), 293--312.


\bibitem{WZ}
E. Witten and B. Zwiebach.
\newblock Algebraic structures and differential geometry in two-%
dimensional string theory.
\newblock Nucl. Phys. B 377 (1992), 55-112.

\bibitem{Z}
B. Zwiebach.
\newblock Closed string field theory: Quantum action and the
Batalin-Vilkovisky master equation.
\newblock Nucl. Phys. B 390 (1993), 33-152.

\end{thebibliography}
\end{document}